\newif\ifdraft
\draftfalse

\RequirePackage{fix-cm}
\ifdraft \documentclass[smallextended,referee]{svjour3}
\else \documentclass[twocolumn]{svjour3} \fi

\usepackage{graphicx}
\usepackage{textcomp}
\usepackage{amsmath,amssymb}
\usepackage[dvipdfm,colorlinks,allcolors=blue]{hyperref}
\ifdraft
\usepackage[nolists,noheads,nomarkers,tablesfirst]{endfloat}
\usepackage{lineno} \linenumbers
\fi

\newcommand{\fake}[1]{\leavevmode\hphantom{#1}}

\journalname{Radiological Physics and Technology}

\begin{document}

\title{Estimation of linear energy transfer distribution for broad-beam carbon-ion radiotherapy at the National Institute of Radiological Sciences, Japan}
%\thanks{Notes about the article}
%\subtitle{Subtitle}

\titlerunning{Estimation of LET for carbon-ion radiotherapy}

\author{Nobuyuki~Kanematsu \and Naruhiro~Matsufuji \and Taku~Inaniwa}
\authorrunning{N.~Kanematsu \and N.~Matsufuji \and T.~Inaniwa}

\institute{
N.~Kanematsu \quad \email{kanematsu.nobuyuki@qst.go.jp}
\at Medical Physics Section, National Institute of Radiological Sciences Hospital, QST; 4-9-1 Anagawa, Inage-ku, Chiba 263-8555, Japan
\and 
N.~Matsufuji \and T.~Inaniwa 
\at Department of Accelerator and Medical Physics, National Institute of Radiological Sciences, QST; 4-9-1 Anagawa, Inage-ku, Chiba 263-8555, Japan
}

\date{\today}

\maketitle

\begin{abstract}
Carbon-ion radiotherapy (CIRT) is generally evaluated with the dose weighted by relative biological effectiveness (RBE), while the radiation quality varying in the body of each patient is ignored for lack of such distribution.
In this study, we attempted to develop a method to estimate linear energy transfer (LET) for a treatment planning system that only handled physical and RBE-weighted doses.
The LET taken from a database of clinical broad beams was related to the RBE per energy with two polyline fitting functions for spread-out Bragg peak (SOBP) and for entrance depths, which would be selected by RBE threshold per energy per modulation.
The LET estimation was consistent with the original calculation typically within a few keV/$\mu$m except for the overkill at the distal end of SOBP.
The CIRT treatments can thus be related to the knowledge obtained in radiobiology experiments that used LET to represent radiation quality.
\keywords{relative biological effectiveness \and linear energy transfer \and carbon-ion radiotherapy \and treatment planning \and retrospective analysis}
\end{abstract}

\section{Introduction}

Use of heavy charged particles for cancer treatment was initiated by historical pioneers \cite{Endo2017}, and is rapidly expanding worldwide \cite{Jermann2015}.
In particular, carbon-ion radiotherapy (CIRT) features finely customizable dose distribution with concurrent enhancement of relative biological effectiveness (RBE), and has been practiced for over 20 years with technological innovations and clinical outcomes \cite{Kamada2015}.
However, despite continuous efforts in experimental radiobiology \cite{Friedrich2012}, the radiobiological knowledge has been applied to CIRT only conceptually per disease \cite{Schlaff2014}.
The application to individual treatments has been limited for the difficulty in quantifying the radiation quality, or spectrum of ionizing particles, varying in the body of each patient.

In CIRT, dose $D$ weighted by RBE $r_\text{BE}$, or RBE-weighted dose $r_\text{BE}\,D$, is commonly used for treatment planning. 
The RBE is generally defined as 
\begin{equation}
r_\text{BE} = \left.\frac{D_\text{ref}}{D}\right|_\text{endpoint},
\end{equation}
where $D_\text{ref}$ is the dose of a reference radiation for the same biological endpoint of a reference cell line as with the current dose.
Among other influential factors for RBE, the radiation quality is generally most focused and is typically represented by linear energy transfer (LET) to tissue per ion track averaged with dose weighting (dose-averaged) \cite{Furusawa2000,Suzuki2000}, or
\begin{equation}
L_\text{D} = \frac{\sum_i L_i D_i}{\sum_i D_i}\approx \frac{\sum_i S_i^2}{\sum_i S_i}, \label{eq:2}
\end{equation}
where $L_i$ is the LET from ion track $i$, $D_i$ is its dose contribution ($\propto L_i$), and $S_i$ is the stopping power of water for the ion \cite{ICRU2014}, or the expectation value of the LET.
While the RBE is the effectiveness of the current dose for reference cells in a reference condition, the LET (dose-averaged, unless otherwise noted hereafter) may potentially give a clue to the sensitivity of various cells in various conditions based on the collection of radiobiology experiments with ion beams \cite{Ando2009,Friedrich2012}.
The spatial distribution of LET for each CIRT plan, if available, will thus allow radiation oncologists to further consider the radiosensitivity of relevant tissue cells \cite{Inaniwa2017}.

Clinical studies on CIRT have been conducted with a broad-beam facility at the National Institute of Radiological Sciences (NIRS) in Japan since 1994 \cite{Kanai1999,Torikoshi2007}, where two systems were mainly used for treatment planning.
The first system HIPLAN (in-house development) employed the broad-beam algorithm ignoring beam blurring except around field edges \cite{Endo1996}. 
The second system XiO-N (Mitsubishi Electric, Tokyo) employed the pencil-beam algorithm for more accurate heterogeneity correction \cite{Kanematsu2011}.
These systems were intrinsically designed to be used with the clinical RBE of original NIRS definition: 
The particle spectrum including projectile fragments was semi-analytically simulated in water \cite{Sihver1998}. 
In the linear quadratic (LQ) modeling of the carbon-ion beams \cite{Kanai1999}, the dose-averaged LET was used as a parameter to relate the dose to the survival of human salivary gland tumor (HSG) cells in radiobiology experiments \cite{Furusawa2000}.
The RBE varying with depth (in water, unless otherwise noted) was calculated for the endpoint of 10\% survival fraction (SF) with rescaling to 3.0 at 80 keV/$\mu$m for the compatibility with historical fast-neutron therapy in dose prescription \cite{Matsufuji2007}.

With the launch of a scanning-beam facility at NIRS in 2011, the clinical RBE was reformulated using the microdosimetric kinetic model with Monte Carlo simulation for accurate handling of secondary particles, for a plan-specific endpoint of variable SF, and against a reference carbon-ion radiation with rescaling for the compatibility of dose prescription \cite{Inaniwa2015}.
Around the same time, an extra energy of $E/A=$ 430 MeV was introduced and all the ridge filters were remanufactured for the broad-beam system to follow the new RBE definition for a slightly modified endpoint of 30\% in HSG-cell SF, or the most common prescription of 3.6 Gy in RBE-weighted dose \cite{Tsujii2012}.
The XiO-N system was set up for the clinical use with the reconfigured beams despite under its intrinsic limitation of the fixed SF regardless of actual dose prescription.

The HIPLAN system could optionally calculate the distributions of physical dose and LET as well as RBE-weighted dose, though the functionality was opted out for more than 7000 patients in the past.
The XiO-N system has automatically stored the physical and RBE-weighted dose distributions for more than 2000 patients so far in preparation for possible verification dosimetry. 
To provide LET distributions for retrospective analysis of the archived XiO-N plans and for ongoing treatments, the obsolete HIPLAN system must be reused for less accurate and troublesome plan reproduction.
In this study, we have attempted to develop an alternative and potentially semiautomatic method to derive LET distributions for the XiO-N plans.

\section{Materials and methods}

\subsection{Clinical carbon-ion beams}

For broad-beam CIRT at NIRS, carbon ions of kinetic energy per nucleon $E/A=$ 290, 350, or 400 MeV have been mainly used with a wobbler-scatterer system for field formation, a selectable ridge filter for range modulation between $M=$ 2 cm and 15 cm, and a variable range shifter so that any tumor can be optimally covered with a spread-out Bragg peak (SOBP) \cite{Kanematsu2007}.
The HIPLAN system was set up with the beam data including physical dose, RBE-weighted dose, and LET, which all are of theoretical calculation as functions of depth.
The XiO-N system was set up with measured physical dose and redefined RBE.

\begin{figure}
\includegraphics{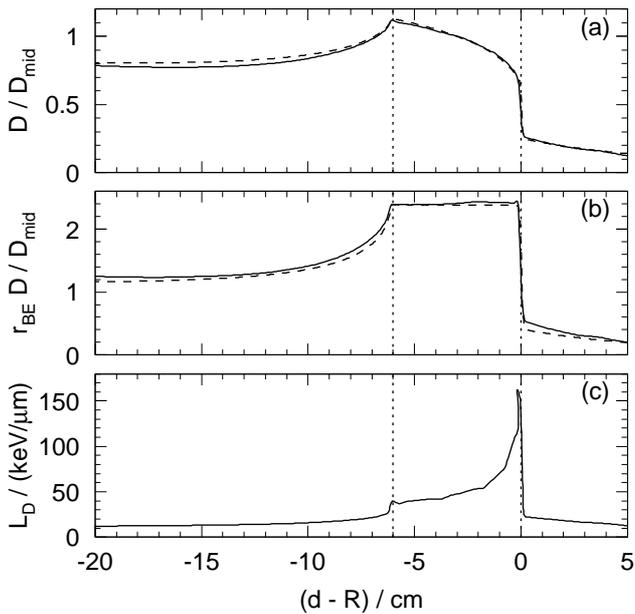}
\caption{(a) Physical dose $D$ and (b) RBE-weighted dose $r_\text{BE}\,D$ relative to mid-SOBP dose $D_\text{mid}$, and (c) LET $L_\text{D}$ as functions of range-subtracted depth $(d-R)$ in water registered in HIPLAN (solid lines) and in XiO-N (dashed lines) for the beam of $E/A=$ 350 MeV and $M=$ 6 cm, and boundaries between the Entrance, SOBP, and Tail depth regions (dotted lines).}
\label{fig:1}
\end{figure}

Figure~\ref{fig:1} shows a typical clinical beam registered in the HIPLAN and XiO-N systems.
The high similarity between the two systems suggests that the differences in physical modeling and in biological endpoint were mostly minor or canceling.
To deal with such modulated beams, we defined three depth regions: $d < R-M$ (Entrance), $R-M \le d \le R$ (SOBP), and $R < d \le R + 5$ cm (Tail), where $d$ is the depth in water and $R$ is the beam range of continuous slowing-down approximation \cite{ICRU2014} and the depths of $d > R + 5$ cm were excluded for the insufficient number of significant digits for the HIPLAN doses.
The RBE of XiO-N was thus related to the LET of HIPLAN at 1-mm depth intervals for each beam.

A portion of primary carbon ions break up into secondary fragments by the nuclear interactions with matter, whose yield may vary with range or initial beam energy.
In the Entrance region, the carbon ions have relatively low LET, and strong positive correlation is theoretically expected between LET and RBE \cite{Kellerer1972}.
In the SOBP region, the locally stopping high-LET carbon ions cause the saturated biological effect, or overkill, and spoil the correlation especially when mixed with low-LET fragments.
In the Tail region, the fragments of wide variety in LET dominate, though with low dose.

\subsection{Formulation of RBE--LET relations}

We formulated LET estimation $\hat{L}_\text{D}$ covering all the clinical beams per energy with manually fitted polyline functions of RBE, where the Entrance and SOBP regions were handled separately to deal with discontinuity at their boundary as implied by Fig.~\ref{fig:1}.
The Tail region was ignored for its similarity to the Entrance region in RBE as well as for its low dose contribution. 
The indifference to the fragments may also be justified by their theoretical difficulty such that even advanced Monte Carlo simulation required empirical correction for them \cite{Sakama2012}.

To select either polyline to apply for a single $\hat{L}_\text{D}$ value, we took the RBE at the proximal end of SOBP, or at $d = R-M$, as the RBE threshold ${r_\text{BE}}_\text{th}$ per energy per modulation.
For a RBE above the threshold, the function for SOBP will be applied, or otherwise the function for Entrance will be applied, to estimate the LET.

\subsection{Application of LET estimation}

\paragraph{LET reproducibility}
We estimated LET from RBE of XiO-N and compared with the LET of HIPLAN for the beam of $E/A=$ 350 MeV and $M=$ 6 cm, where the beam range was virtually adjusted to set the SOBP center position at the isocenter in water.

\paragraph{Radiation mixing}
Similarly, we simulated an opposing arrangement of the same beams to examine the radiation-mixing effect. 
We first estimated LET from RBE of XiO-N for the individual beams, then applied dose-averaging over the beams, and compared the resultant LET estimation with the dose-averaged LET of HIPLAN.

\section{Results}

\subsection{Formulation of RBE--LET relations}

\begin{figure}
\includegraphics{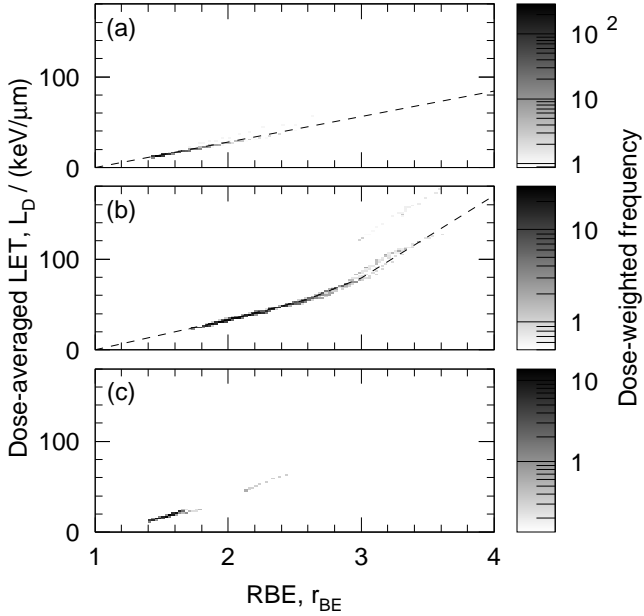}
\caption{Dose-weighted frequency distributions of LET and RBE of XiO-N, accumulated over all modulated beams of $E/A=$ 350 MeV for 1-mm-interval depths of (a) Entrance and (b) SOBP with polyline fitting functions (dashed lines), and of (c) Tail.}
\label{fig:2}
\end{figure}

Figure \ref{fig:2} shows the correlation between LET and RBE for $E/A=$ 350 MeV with the polyline fitting functions, where we ignored the entries around $r_\text{BE} \gtrsim$ 3.0 and $L_\text{D} \gtrsim$ 100 keV/$\mu$m at the distal end of SOBP, where severe overkill should occur, to make the function as single-valued. 
Table \ref{tab:1} enumerates the polyline fitting results for all the HIPLAN energies, where a single linear relation was commonly applicable to Entrance as expected while the overkill effect slightly increased with energy for SOBP.

\begin{figure}
\includegraphics{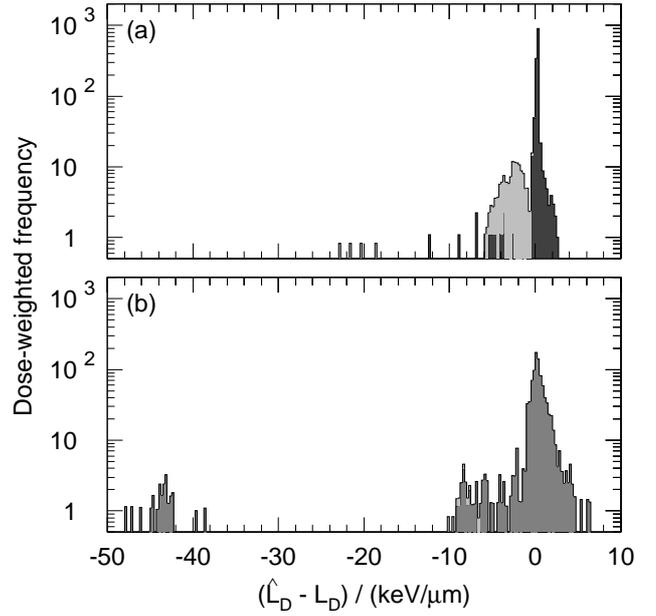}
\caption{Dose-weighted frequency distributions of residual deviation between the LET $\hat{L}_\text{D}$ estimated from RBE of XiO-N and the LET $L_\text{D}$ of HIPLAN: (a) for $r_\text{BE} < {r_\text{BE}}_\text{th}$ and (b) for $r_\text{BE} \ge {r_\text{BE}}_\text{th}$, at 1-mm-interval depths of all modulated beams of $E/A=$ 350 MeV. 
The histogram entries were constituted by Entrance (dark gray), SOBP (medium gray), and Tail (light gray).}
\label{fig:3}
\end{figure}

Table \ref{tab:2} shows the RBE at the proximal end of SOBP, or RBE threshold ${r_\text{BE}}_\text{th}$, per energy per modulation.
For $E/A=$ 430 MeV, which was not supported by HIPLAN, the fitting functions for 400 MeV should be applied as the best alternative.
Figure \ref{fig:3} shows the distributions of residual deviation between the LET estimated from RBE of XiO-N and the LET of HIPLAN for $E/A=$ 350 MeV, where the RBE threshold separated the SOBP from the other regions effectively and the (mean $\pm$ standard deviation) amounted to $(-0.01 \pm 1.30)$ keV/$\mu$m for $r_\text{BE} < {r_\text{BE}}_\text{th}$ and $(-0.60 \pm 5.98)$ keV/$\mu$m for $r_\text{BE} \ge {r_\text{BE}}_\text{th}$.

\begin{table}
\caption{RBE--LET relations $(r_\text{BE}, \hat{L}/u)$ with $u =$ 1 keV/$\mu$m for XiO-N, to constitute the polyline fitting functions for Entrance and for SOBP of $E/A=$ 290, 350, and 430 MeV.}
\label{tab:1}
\begin{tabular}{ccccc}
\hline\hline\noalign{\smallskip}
& \multicolumn{4}{c}{$(r_\text{BE}, \hat{L}/u)$} \\
\noalign{\smallskip}\cline{2-5}\noalign{\smallskip}
& & \multicolumn{3}{c}{SOBP} \\
\noalign{\smallskip}\cline{3-5}\noalign{\smallskip}
\textnumero & Entrance & 290 MeV & 350 MeV & 400 MeV \\
\noalign{\smallskip}\hline\noalign{\smallskip}
1 & $(1.0,\fake{0}0)$ & $(1.0,\fake{00}0)$ & $(1.0,\fake{00}0)$ & $(1.0,\fake{00}0)$ \\
2 & $(4.0,84)$ & $(2.0,\fake{0}31)$ & $(2.0,\fake{0}33)$ & $(2.0,\fake{0}33)$ \\
3 & & $(2.6,\fake{0}55)$ & $(2.6,\fake{0}56)$ & $(2.6,\fake{0}56)$ \\
4 & & $(3.0,\fake{0}72)$ & $(3.0,\fake{0}79)$ & $(3.0,\fake{0}85)$ \\
5 & & $(4.0,140)$ & $(4.0,170)$ & $(4.0,180)$ \\
\noalign{\smallskip}\hline\hline
\end{tabular}
\end{table}

\begin{table}
\caption{RBE thresholds ${r_\text{BE}}_\text{th}$ for $E/A=$ 290, 350, 400, and 430 MeV per range modulation $M$, to select the function either for Entrance ($r_\text{BE} < {r_\text{BE}}_\text{th}$) or for SOBP ($r_\text{BE} \ge {r_\text{BE}}_\text{th}$) to estimate LET for XiO-N.}
\label{tab:2}
\begin{tabular}{ccccc}
\hline\hline\noalign{\smallskip}
 & \multicolumn{4}{c}{${r_\text{BE}}_\text{th}$} \\
\noalign{\smallskip}\cline{2-5}\noalign{\smallskip}
$M$/cm & 290 MeV & 350 MeV & 400 MeV & 430 MeV \\
\hline\noalign{\smallskip}
\fake{0}2.0 & 2.769 & 2.690 & 2.619 & 2.563 \\
\fake{0}2.5 & 2.632 & 2.565 & 2.498 & 2.447 \\
\fake{0}3.0 & 2.525 & 2.466 & 2.404 & 2.358 \\
\fake{0}4.0 & 2.378 & 2.314 & 2.254 & 2.213 \\
\fake{0}5.0 & 2.263 & 2.203 & 2.147 & 2.117 \\
\fake{0}6.0 & 2.176 & 2.116 & 2.060 & 2.021 \\
\fake{0}7.0 & 2.094 & 2.045 & 1.989 & 1.957 \\
\fake{0}8.0 & 2.040 & 1.982 & 1.935 & 1.903 \\
\fake{0}9.0 & 1.984 & 1.932 & 1.882 & 1.853 \\
10.0 & 1.941 & 1.888 & 1.842 & 1.812 \\ 
11.0 & 1.900 & 1.848 & 1.805 & 1.775 \\
12.0 & 1.860 & 1.814 & 1.771 & 1.742 \\
15.0 & --- & 1.726 & 1.686 & 1.660 \\
\noalign{\smallskip}\hline\hline
\end{tabular}
\end{table}

\subsection{Application of LET estimation}

\begin{figure}
\includegraphics{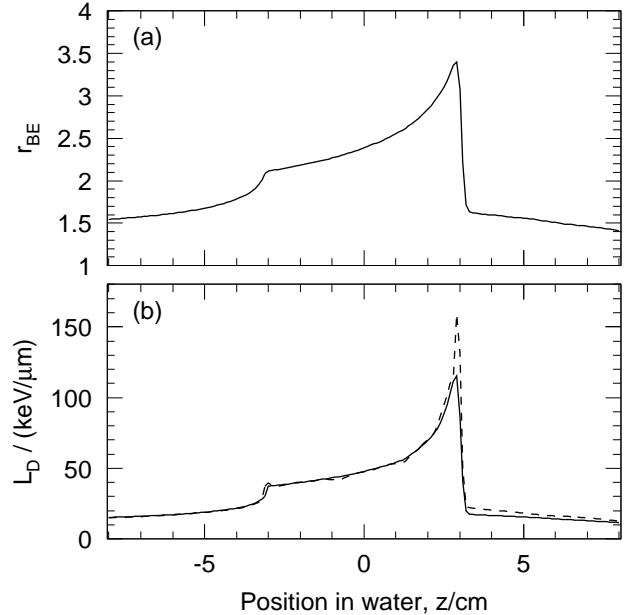}
\caption{Distributions of (a) RBE of XiO-N and (b) LETs of XiO-N estimation $\hat{L}_\text{D}$ (solid) and of HIPLAN $L_\text{D}$ (dashed) along the beam central axis in water for a single beam of $E/A=$ 350 MeV and $M=$ 6 cm.}
\label{fig:4}
\end{figure}

\paragraph{LET reproducibility}
Figure \ref{fig:4} shows the RBE and LET distributions in the single-beam case.
The LET estimation generally agreed well with the LET of HIPLAN except for the peak at the distal end of SOBP and for Tail, as it was designed as such. 
The peak LET of 157 keV/$\mu$m was underestimated by 26\%.

\begin{figure}
\includegraphics{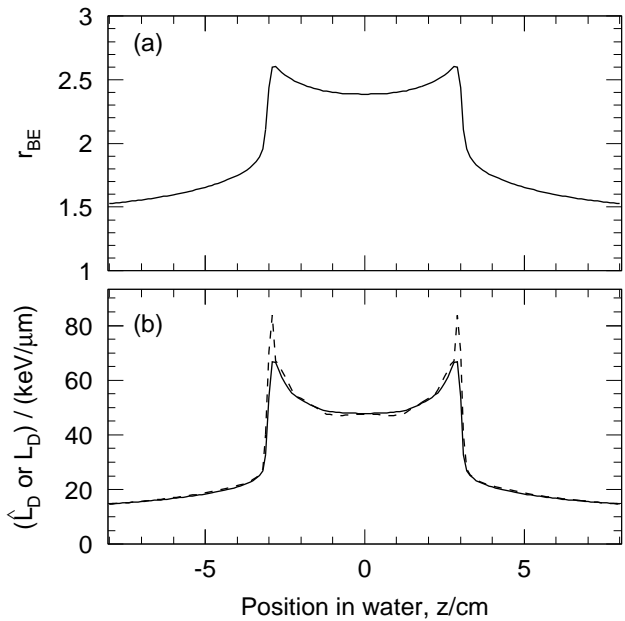}
\caption{Distributions of (a) RBE of XiO-N and (b) LETs of XiO-N estimation $\hat{L}_\text{D}$ (solid) and of HIPLAN $L_\text{D}$ (dashed) along the beam central axis in water for opposing beams of $E/A=$ 350 MeV and $M=$ 6 cm.}
\label{fig:5}
\end{figure}

\paragraph{Radiation mixing} 
Figure \ref{fig:5} shows the dose-averaged RBE and LET distributions in the opposing-beam case.
Even with radiation mixing, the LET estimation generally agreed well with the LET of HIPLAN.
The disagreement observed for Tail of the single beam was mitigated by dose averaging with Entrance of another beam, while the peak LET of 84 keV/$\mu$m was still underestimated by 20\%. 

\section{Discussion}

The present LET estimation was formulated for the RBE that was defined for a fixed SF, or independently of dose, though conceptually against the LQ model.
With dose-dependent RBE, the dose must also be considered for the LET estimation.
The LET data excerpted from HIPLAN only covered primary carbon ions and projectile fragments, which may be acceptable for the following reasons.
The target nuclear fragments and recoils are highly ionizing and short ranged, and thus difficult to calculate theoretically or to measure experimentally.
In addition, the LET of the same definition has been commonly used to represent radiation quality in radiobiology experiments as well.

For a LET as high as to cause overkill, single-valued LET estimation from RBE is intrinsically limited.
Hopefully, that will not be clinically critical because such a overkill region is likely in the margin against the potential spread of cancer cells around a tumor. 
Conversely, the insensitivity to overkill may be an intrinsic deficiency of LET for a clinical index of radiation quality, and could confuse treatment analysis, especially with dose averaging over multiple beams.
In this regard, the direct use of RBE distributions may offer better correlation with the clinical results because the overkill effect has been corrected for the HSG cells which were assumed to represent all cells.

The estimation of LET from RBE was demonstrated to work for the sample beam, and should naturally work for the other clinical broad beams of NIRS as well.
The correlation between RBE and LET was high as illustrated in Fig.~\ref{fig:2}, which was also true for the other energies, and the variation of LET--RBE relation among energies was rather marginal as shown in Table \ref{tab:1}. 
This fact may also justify the application of the fitting function for 400 MeV to 430 MeV.
Because the RBE in Entrance and Tail should always be lower than the RBE threshold, or the minimum RBE in SOBP, the appropriate function can be selected reliably by the RBE comparison according to Table \ref{tab:2}.
The present LET estimation was intended for the broad beams of CIRT at NIRS, and may approximately apply to similar beams at other CIRT facilities. 

\section{Conclusions}

For the CIRT treatments of over 2000 patients planned with the XiO-N system, we have developed a method to estimate their LET distributions from the RBE distributions available from the plans. 
The estimated LET was consistent with the original calculation typically within a few keV/$\mu$m in deviation except for the overkill at the distal end of SOBP.
The LET estimation to relate radiation quality of CIRT to that of radiobiology experiments will facilitate the radiobiological interpretation of the treatments retrospectively or prospectively in treatment planning.

\begin{acknowledgements}
The authors are deeply indebted to all the researchers, medical staff, and technical staff of NIRS who have prepared, practiced, or supported the extensive clinical studies on carbon-ion radiotherapy over many years.
\\
\noindent\textbf{Conflict of interest}
The authors declare that they have no conflict of interest.
\\
\noindent\textbf{Ethical approval}
No ethical approval was required for this article as it does not contain any studies with human subjects or animals. 
\end{acknowledgements}

\bibliographystyle{vancouver}
\bibliography{references}

\end{document}